\documentclass[bimj,fleqn]{w-art}
\usepackage{times}
\usepackage{w-thm}
\usepackage[authoryear]{natbib}
\usepackage{rotating}

\theoremstyle{plain}

\theoremstyle{definition}

\usepackage[]{graphicx}
\usepackage[graphicx]{realboxes}
\chardef\bslash=`\\ 

\begin{document}
\DOIsuffix{bimj.200100000} \Volume{XX} \Issue{XX} \Year{2021}
\pagespan{1}{}
\keywords{Propensity score; Standardized difference; Weighted z-difference; Z-difference;\\
}

\title[The z-difference in weighted propensity score analyses]{Measuring covariate balance in weighted propensity score analyses by the weighted z-difference}
\author[Filla {\it{et al.}}]{Tim Filla\inst{1}}
\address[\inst{1}]{Department of Medical Biometry and Bioinformatics, Heinrich Heine University D{\"u}sseldorf, Moorenstr. 5, 40225 D{\"u}sseldorf, Germany}
\author[]{Holger Schwender\inst{2}}
\address[\inst{2}]{Mathematical Institute, Heinrich Heine University D{\"u}sseldorf, Universit{\"a}tsstr. 1, 40225 D{\"u}sseldorf, Germany}
\author[]{Oliver Kuss\footnote{Corresponding author: {\sf{e-mail: oliver.kuss@ddz.de}}, Phone: +00-999-999-999, Fax: +00-999-999-999}\inst{3}}
\address[\inst{3}]{Institute for Biometrics and Epidemiology, German Diabetes Center, Leibniz Institute for Diabetes Research at Heinrich Heine University D{\"u}sseldorf, Auf'm Hennekamp 65, 40225 D{\"u}sseldorf, Germany}

\Receiveddate{zzz} \Reviseddate{zzz} \Accepteddate{zzz}

\begin{abstract}
Propensity score (PS) methods have been increasingly used in
recent years when assessing treatment effects in nonrandomized
studies. In terms of statistical methods, a number of new PS
weighting methods were developed, and it was shown that they can
outperform PS matching in efficiency of treatment effect
estimation in different simulation settings. For assessing balance
of covariates in treatment groups, PS weighting methods commonly
use the weighted standardized difference, despite some
deficiencies of this measure like, for example, the distribution
of the weighted standardized difference depending on the sample
size and on the distribution of weights. We introduce the weighted
z-difference as a balance measure in PS weighting analyses and
demonstrate its usage in a simulation study and by applying it to
an example from cardiac surgery. The weighted z-difference is
computationally simple and can be calculated for continuous,
binary, ordinal and nominal covariates. By using Q-Q-plots we can
compare the balance of PS weighted samples immediately to the
balance in perfectly matched PS samples and to the expected
balance in a randomized trial.
\end{abstract}

\maketitle                   






\section{Introduction}
Propensity score (PS) methods are widely used to control for
confounding in nonrandomized trials, and have several
epistemological as well as statistical advantages as compared to
standard regression modelling (\cite{kuss2016propensity}). In
essence, propensity score modelling consists of a two-step
procedure. In a first step, the individual propensity score, the
probability of being treated given the covariates, is estimated,
usually by logistic regression. In the second step, the treatment
effect is estimated preferably by using either weighting for the
PS or matching on the PS (\cite{austin1,austin2}). When using PS
weighting, inverse probability treatment weights
(IPTW, 	\cite{robins2000marginal}) are generally used, but also
several new methods, e.g., matching weights (\cite{li2013weighting})
or overlap weights (\cite{RN1025}) have been developed.

An important part of each PS analysis is to assess balance of
covariates between treatment groups in the analysis
sample (\cite{bib7}). To this task, a number of statistical measures
has been introduced (\cite{belitser2011measuring,franklin2014metrics}), of which the
standardized difference (\cite{austin2009using}) is used most widely.
For matched PS analysis, \cite{RN993} has advanced the idea of
\cite{bib10} and \cite{bib11} and proposed the
z-difference as a balance measure. The z-difference measures
balance on a scale that is standard Gaussian under the null
hypothesis of balance, that is, if there is no difference in
baseline distributions between treatment and control group. As
compared to the standardized difference, the z-difference has the
two main advantages that its distribution does not depend on
sample size and it allows comparing balance of baseline covariates
on all scales, that is, for continuous, binary, ordinal, or
nominal covariates (\cite{RN993}). In PS matching, the z-difference
has been successfully used in applied research by
us (\cite{bib12,bib13}) and others (\cite{bib14,bib15,bib16}).

However, the z-difference cannot be used with PS weighting,
because it can not deal with the individual PS weights that are
attached to each observation, but it would be certainly useful to
have balance measures also for weighted PS analyses. \cite{austinweighted}
proposed a weighted standardized difference to this task, but this
inherits the disadvantages of the unweighted standardized
difference. That is, it is only defined for
continuous and binary variables, and, as we will show later, its
distribution also depends on the sample size. Moreover, and as an
additional disadvantage, the distribution of the weighted
standardized difference depends on the distribution of the weights
themselves, which additionally impedes its interpretation.

In this article, we propose a weighted z-difference that does not
have these disadvantages and can be used to assess covariate
balance in weighted PS analyses while inheriting the advantages of
the original z-difference. In section \ref{methods} we re-iterate
the unweighted z-difference and explain the difference to the
unweighted standardized difference. In section
\ref{chapterweightedzdiff} the weighted z-difference is introduced
for the four different possible scales of covariates, that is,
continuous, binary, ordinal and nominal. Section
\ref{simulation_chapter} gives results from simulations that
demonstrate the advantages of the weighted z-difference in
comparison to the weighted standardized difference. In section
\ref{real_example} the weighted z-difference is applied to a
published PS matching analysis using IPTW and matching weights. We
finish with a discussion of our findings in section
\ref{discussion}.

\section{The z-difference and the standardized difference}\label{methods}
For notation, we assume that a data set with $N$ observations and
$P$ covariates is available, where we aim for these covariates to
be balanced in our PS analysis. In addition, a binary treatment
variable, whose effect estimate is our primary interest, is given.
Before final effect estimation, the balance of covariates in the
two treatment groups should be assessed, and the z-difference as
well as the standardized difference have been proposed to this
task.

For re-iterating the idea of the original z-difference we consider
a single covariate, where we denote the $n$ observations from the
treatment group as $x_1,\dots,x_n$ and the $m:=N-n$ observations
from the control group as $y_1,...,y_m$. Throughout the paper we
will make the assumption that the variables $X_1,\dots,X_n$ and
$Y_1,\dots,Y_m$ are independently and identically distributed
$(i.i.d.)$ with potentially different distributions in the two
groups. Expected values and variances of the two distributions are
denoted by E$(X_i) = \mu_X,\ $Var$(X_i) = \sigma^2_X$, and $E(Y_j)
= \mu_Y,\ $Var$(Y_j)=\sigma^2_Y$. We estimate $\mu_X$ by
$\hat{\mu}_X:=\bar{X}= \sum_{i=1}^n\dfrac{X_i}{n}$ and
$\sigma^2_X$ by $\hat{\sigma}_X^2:=\sum_{i=1}^n\dfrac{\left(X_i -
\bar{X}\right)^2}{n-1}$, and analogously for $\mu_Y$ and
$\sigma^2_Y$.

The basic idea of the original z-difference was to measure
covariate balance by a statistic that is asymptotically standard
Gaussian distributed under the null hypothesis that some
properties of the two distributions of $X_i$ and $Y_j$ are equal.
For example, in the case of a continuous covariate the null
hypothesis might be that $\mu_X = \mu_Y$, or
$\sigma^2_X=\sigma^2_Y$, or both. To be concrete, to test the null
hypothesis $\mu_X = \mu_Y$ we previously used the mean difference
$\hat{\mu}_{X}-\hat{\mu}_{Y}$ to measure distance of means between
groups. By dividing the mean difference by its standard deviation
we achieve the z-difference
\begin{equation}\label{zdiff}
\mbox{$Z_{c}$}:=\dfrac{\hat{\mu}_{X}
-\hat{\mu}_{Y}}{\sqrt{\dfrac{\hat{\sigma}_X^2}{n} +
\dfrac{\hat{\sigma}_Y^2}{m}}},
\end{equation}
and as a consequence of the central limit theorem this
z-difference is asymptotically standard Gaussian distributed under
the null hypothesis.

Opposed to the z-difference the standardized difference for
continuous covariates is defined by \cite{austin2009using}
\begin{equation}\label{stddiff}
\mbox{$Sd_{c}$}:=\dfrac{\hat{\mu}_{x} -
\hat{\mu}_{y}}{\sqrt{\dfrac{\hat{\sigma}_X^2 +
\hat{\sigma}_Y^2}{2}}}.
\end{equation}

If we additionally assume $\sigma_X^2=\sigma_Y^2=:\sigma^2$ then
we can replace $\hat{\sigma}_X^2$ and $\hat{\sigma}_Y^2$ by a
pooled variance estimator $\hat{\sigma}_p^2$ and the formulas of
the z-difference and standardized difference simplify to
\begin{equation}
\mbox{$Z_{c}$}=\dfrac{\hat{\mu}_{X}
-\hat{\mu}_{Y}}{\sqrt{\hat{\sigma}_p^2\left(\dfrac{1}{n} +
\dfrac{1}{m}\right)}},
\end{equation}
and
\begin{equation}
\mbox{$Sd_{c}$}=\dfrac{\hat{\mu}_{x} -
\hat{\mu}_{y}}{\sqrt{\hat{\sigma}_p^2}}.
\end{equation}
Comparing z-difference and standardized difference we find $Z_c =
Sd_c/\sqrt{{\frac{1}{n}+\frac{1}{m}}}$. Remembering that the
z-difference is $\mathcal{N}(0,1)$ distributed under the null
hypothesis, it follows that the standardized difference is
distributed $\mathcal{N}\left(0,{\frac{1}{n}+\frac{1}{m}}\right)$
(\cite{austin2015moving}) and thus depends on the sample size.


\section{The weighted z-difference}\label{chapterweightedzdiff}

In weighted PS analyses, each single observation comes with a PS
weight $w_{X_i}>0,\ w_{Y_j}>0,\ i=1,\dots,n,\ j=1,\dots,m$, that
is, an additional multiplicative factor. For computational reasons
we assume that the weights in each group are scaled, that is,
$\sum_{i=1}^nw_{X_i} = \sum_{j=1}^mw_{Y_j} =1$. As the original
standard z-difference was only defined for matched PS analysis
where no weights are given, we introduce the weighted z-difference
in the following. It will be seen that the weighted z-difference
generalizes the z-difference, where the z-difference results when
$w_{X_1}=\dots= w_{X_n}$ and $w_{Y_1}=\dots=w_{Y_m}$. All weighted
z-differences are implemented in the R package weightedZdiff
(\cite{filla}), which is available on the Comprehensive R Archive
Network (CRAN).

\subsection{The continuous case: Comparing means}
For continuous covariates we propose to compare the expected mean
values $\mu_{wX},\mu_{wY}$ in the weighted sample by their
difference to check the null hypothesis $\mu_{wX}=\mu_{wY}$. The
expected value in a weighted sample is estimated by
$\hat{\mu}_{wX} = \bar{X}_w = \sum\limits_{i=1}^{n}w_{X_i}X_i$,
and analogously for $\mu_{wY}$. Calculation of the standard
deviation for the weighted mean difference is straightforward
using the independence assumption for the variables $X_i$ and
$Y_j$. We obtain
\begin{equation}
Var(\hat{\mu}_{wX}) = \sum_{i=1}^nw_{X_i}^2\sigma_X^2,
\end{equation}
and the analogous result for Var$(\hat{\mu}_{wY})$.

Finally by using the weighted variance estimator
(\cite{austin2015moving})
\begin{equation}\label{gleichungvarianz}
\hat{\sigma}_{wX}^2  =
\dfrac{1}{1-\sum_{i=1}^nw_{X_i}^2}\sum_{i=1}^nw_{X_i}\left(X_i-\bar{X}_w\right)^2
\end{equation} for $\sigma^2_{X}$, and analogously for $\sigma^2_{Y}$, the weighted
z-difference for comparing means of continuous covariates is
defined by
\begin{equation}\label{weighted_continuous_zdiff}
Z_{w,c}:=\dfrac{\hat{\mu}_{wX} -
\hat{\mu}_{wY}}{\sqrt{\hat{\sigma}^2_{wX}\sum\limits_{i=1}^{n}w_{X_i}^2
+\hat{\sigma}^2_{wY}\sum\limits_{j=1}^{m}w_{Y_j}^2}}.
\end{equation}

The standard Gaussian distribution of $Z_{w,c}$ under the null
hypothesis of balance follows from the central limit theorem for
weighted observations of Lindeberg (\cite{chowundteicher}).


The relation of the z-difference and the standardized difference
carries over to the weighted case. Under the assumption of equal
variances $\sigma_{wX}^2 = \sigma_{wY}^2$ we can use a pooled
variance estimator $\sigma_{wp}^2$ and the formula of $Z_{w,c}$
reduces to
\begin{equation}
Z_{w,c}=\dfrac{\hat{\mu}_{wX} -
\hat{\mu}_{wY}}{\sqrt{\left(\sum\limits_{i=1}^{n}w_{X_i}^2
+\sum\limits_{j=1}^{m}w_{Y_j}^2\right)\hat{\sigma}^2_{wp}}}.
\end{equation}

In parallel, the weighted standardized difference is given by
\begin{equation}
Sd_{w,c}=\dfrac{\hat{\mu}_{wX} -
\hat{\mu}_{wY}}{\sqrt{\hat{\sigma}^2_{wp}}},
\end{equation}
and we find the ratio $Z_{w,c} = Sd_{w,c} \times \left [
\sqrt{\sum\limits_{i=1}^{n}w_{X_i}^2+\sum\limits_{j=1}^{m}w_{Y_j}^2}
\right ]^{-1}$.

As $Z_{w,c}$ is standard Gaussian distributed, $Sd_{w,c}$ follows
a
$\mathcal{N}\left(0,\sum\limits_{i=1}^{n}w_{X_i}^2+\sum\limits_{j=1}^{m}w_{Y_j}^2\right)$
distribution. That is, the distribution of the weighted
standardized difference additionally depends on the distribution
of the weights. In chapter \ref{simulation_chapter} we will show
how this complicates checking covariate balance using the weighted
standardized difference.

\subsection{The continuous case: Comparing variances}
Continuous covariates can not only be different with respect to
their means, but also with respect to higher moments
(\cite{austin2008report,rubin2001using}). Thus we also propose a
weighted z-difference using the difference of weighted variances
$\sigma_{wX}^2 - \sigma_{wY}^2$ for assessing balance, which tests
the null hypothesis $\sigma_{wX}^2 = \sigma_{wY}^2$. For
computational simplicity we go over to the centered variables
$\tilde{X}_i := X_i-\mu_{wX}$ and $\tilde{Y}_i := X_i-\mu_{wY}$ ,
whereby $\mu_{wX}$ and $\mu_{wY}$ are replaced by their weighted
estimators $\bar{X}_w$ and $\bar{Y}_w$. Centering changes the
variance estimator in (\ref{gleichungvarianz}) to
\begin{equation}
\hat{\sigma}_{wX}^2 =
\dfrac{1}{1-\sum_{i=1}^nw_{X_i}^2}\sum_{i=1}^nw_{X_i}\left(X_i-\bar{X}_w\right)^2
=
\dfrac{1}{1-\sum_{i=1}^nw_{X_i}^2}\sum_{i=1}^nw_{X_i}\tilde{X}_i^2
.
\end{equation}

The variance of $\hat{\sigma}_{wX}^2$ is given by
\begin{align*}
Var(\hat{\sigma}_{wX}^2) =&
\dfrac{1}{\left(1-\sum_{i=1}^nw_{X_i}^2\right)^2}\sum_{i=1}^nw_{X_i}^2Var(\tilde{X}^2).
\end{align*}

Finally, if we let $\hat{\sigma}_{w\tilde{X}^2}^2$ be the weighted
variance estimator of Var$(\tilde{X}^2)$, then the weighted
z-difference for the variances of a continuous variable is given
by
\begin{equation}\label{variance_weighted_variance}
 Z_{w,cVar}:=\dfrac{\hat{\sigma}^2_{wX} - \hat{\sigma}^2_{wY}}{\sqrt{\dfrac{\sum_{i=1}^nw_{X_i}^2}{\left(1-\sum_{i=1}^nw_{X_i}^2\right)^2}\hat{\sigma}_{w\tilde{X}^2}^2+\dfrac{\sum_{j=1}^mw_{Y_j}^2}{\left(1-\sum_{j=1}^mw_{Y_j}^2\right)^2}\hat{\sigma}_{w\tilde{Y}^2}^2}}.
\end{equation}

Analogous to the case of means for continuous variables, the
standard Gaussian distribution for $Z_{w,cVar}$ follows from the
central limit theorem for weighted observations and the null
hypothesis $\sigma_{wX}^2 = \sigma_{wY}^2$.


\subsection{The binary case}\label{binarycase}
For binary covariates we use the difference of the weighted
outcome prevalences $p_{wX}$ and $p_{wY}$ as the balance measure.
The respective estimate of $p_{wX}$ in the weighted sample is
$\hat{p}_{wX} = \sum\limits_{i=1}^{n}w_{X_i}X_i$, and analogously
for $p_{wY}$.


Calculation of the variance of $\hat{p}_{wX}$ is straightforward
due to the independence assumption of $X_i$ and $Y_j$ and we find
\begin{equation}
Var(\hat{p}_{wX}) = \sum_{i=1}^nw_{X_i}^2Var(X_i) =
\sum_{i=1}^nw_{X_i}^2p_{X}(1-p_{X}),
\end{equation}
and analogously for Var$(\hat{p}_{wY})$.

Finally we end up with the weighted z-difference for binary
covariates
\begin{equation}\label{weighted_binary_zdiff}
Z_{w,b}:=
\dfrac{\hat{p}_{wX}-\hat{p}_{wY}}{\sqrt{\sum\limits_{i=1}^{n}w_{X_i}^2\hat{p}_{X}(1-\hat{p}_{X})
+\sum\limits_{j=1}^{m}w_{Y_j}^2\hat{p}_{Y}(1-\hat{p}_{Y})}}.
\end{equation}

Again, the standard Gaussian distribution for $Z_{w,b}$ follows
from the central limit theorem for weighted observations and the
null hypothesis $p_{wX} = p_{wY}$.

\subsection{The ordinal case}
For ordinal covariates, the z-difference has to respect the order
of observations and we therefore use the expected mean rank
difference $\mu_{wR_X} - \mu_{wR_Y}$ for ranks $R_{X_i},R_{Y_j},
i=1,\dots,n, j=1,\dots, m$. It is important to note that ranking
is performed only with respect to the observations itself, but
independent of the weights. We estimate the expected mean rank in
the weighted sample by $\hat{\mu}_{wR_X} :=
\sum_{i=1}^nw_{X_i}R_{X_i}$ and $\hat{\mu}_{wR_Y} :=
\sum_{j=1}^mw_{Y_j}R_{Y_j}$.


The variance of the estimated weighted rank difference is given by
\begin{align}\label{eq1}
&Var\left(\hat{\mu}_{wR_X} - \hat{\mu}_{wR_Y}\right) =
Var\left(\sum_{i=1}^{n} w_{X_i}R_{X_i} - \sum_{j=1}^{m}
w_{Y_j}R_{Y_j}\right)=
\sum_{i=1}^{n}w_{X_i}^2Var\left(R_{X_i}\right)+ \sum_{j=1}^{m} w_{Y_j}^2Var\left(R_{Y_j}\right)\notag \\
&+
\left(1-\sum_{i=1}^{n}w_{X_i}^2\right)Cov\left(R_{X_i},R_{X_i'}\right)
+
\left(1-\sum_{j=1}^{m}w_{Y_j}^2\right)Cov\left(R_{Y_j},R_{Y_j'}\right)
- 2Cov\left(R_{X_i},R_{Y_j}\right).
\end{align}

Under the null hypothesis of identically distributed variables
$X_i$ and $Y_j$ we have $Var(R_{X_i})=Var(R_{Y_j})=:\sigma^2_R$,
$Cov(R_{X_i},R_{X_i'}) = Cov(R_{Y_j},R_{Y_j'})$, and
$Cov(R_{X_i},R_{Y_j})=:\sigma_{ij,R}$.


This yields
\begin{eqnarray*}
Var\left(\hat{\mu}_{wR_{X}} - \hat{\mu}_{wR_{Y}}\right)=
\left(\sum_{i=1}^{n}w_{X_i}^2+\sum_{j=1}^{m}w_{Y_j}^2\right)\left(\sigma^2_R-\sigma_{ij,R}\right).
\end{eqnarray*}

Finally we replace $\sigma^2_R$ and $\sigma_{ij,R}$ by their
estimators $\hat{\sigma}^2_{R}$ and $\hat{\sigma}_{ij,R}$ and the
weighted z-difference for ordinal covariates is defined by
\begin{equation}
Z_{w,o} :=\frac{\hat{\mu}_{R_X}-
\hat{\mu}_{R_Y}}{\sqrt{\left(\sum_{i=1}^{n}w_{X_i}^2+\sum_{j=1}^{m}w_{Y_j}^2\right)\left(\hat{\sigma}^2_R-\hat{\sigma}_{ij,R}\right)}}.
\end{equation}

The standard Gaussian distribution of $Z_{w,o}$ follows from the
central limit theorem for weighted observations together with the
theorem of \cite{hajek1968asymptotic} which assures that a
linear function of the ranks converges to a Gaussian distribution.

\subsection{The nominal case}
The derivation of the weighted z-difference for a nominal
covariate relies on the idea of \cite{RN1004} to
compare two histograms of weighted observations. We consider a
nominal covariate with $K$ different categories ($k=1,\dots,K$)
with probabilities $p_{X_k}$ and $p_{Y_k}$ of belonging to
category $k$ of the treatment or control group. We estimate
$p_{X_k}$ by $\hat{p}_{wX_k}:= \sum_{i=1}^nw_i1_{X_i=k}$ and
analogously for $\hat{p}_{wY_k}$, where $1_{X_i=k}$ denotes the
indicator function. Gagunashvili gave an estimator of the variance
of $p_{wX_k}$ by $\hat{\sigma}^2_{w_{Xk}}:=
\sum_{i=1}^nw_{X_i}^21_{X_i=k}$, which is just the sum of squared
weights for each cell of the underlying $2 \times K$ table.
Estimating all $p_{wX_k}$,\ $p_{wY_k}$ and comparing observed and
expected probabilities yields

\begin{equation}
\chi^2_{Z_{w,n}} :=\sum_{k=1}^{K}\frac{\left(\hat{p}_{wX_k}-
\hat{p}_{wY_k}\right)^2}{\hat{\sigma}^2_{w_{Xk}}+\hat{\sigma}^2_{w_{Yk}}}.
\end{equation}

Gagunashvili showed that $\chi^2_{Z_{w,n}}$ is
$\chi^2$-distributed with $K-1$ degrees of freedom under the null
hypothesis of $p_{X_k} = p_{Y_k} = p_k$ for all $k$. We note that
the Gagunashvili statistic closely resembles the standard weighted
$\chi^2$-statistic that we would expect here.
Indeed, the numerator of the Gagunashvili statistic and the
standard weighted $\chi^2$ statistic are identical, however, in
the denominator, Gagunashvili additionally accounts for the
variability of weights.

As the z-difference for nominal covariates should also be standard
Gaussian distributed, we apply a probability integral
transformation (\cite{bib18}) to $\chi^2_{Z_{w,n}}$ and finally
arrive at
\begin{equation}
Z_{w,n} := \Phi^{-1}\left(F(\chi^2_{Z_{w,n}},K-1)\right),
\end{equation}
with $F(\chi^2_{Z_{w,n}},K-1)$ being the cumulative distribution
function $F$ of a $\chi^2$ distribution with $K-1$ degrees of
freedom and $\Phi^{-1}$ the inverse cumulative distribution
function (or quantile) function of the standard normal
distribution.

\section{Simulation study }\label{simulation_chapter}
To assess the statistical properties of the weighted z-difference
in finite samples, we performed three simulation experiments.
First we varied the sample sizes, second we used different methods
for generating the weights and lastly we changed the ratio of
sample sizes in treatment versus control group. In each experiment
we compared the weighted z-difference to the weighted standardized
difference, the current standard for assessing covariate balance
in weighted PS analyses.

\subsection{Continuous covariate, varying sample sizes, uniform weights}
\begin{figure}
\begin{center}
\includegraphics[scale=0.4]{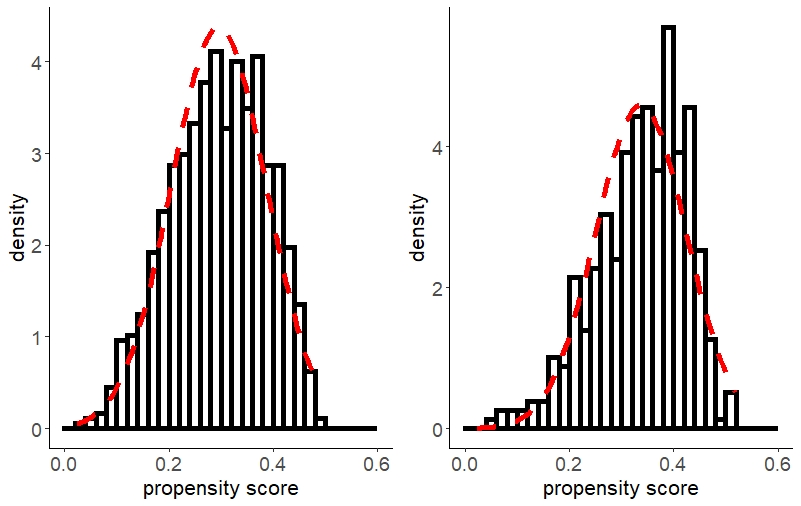}
\caption{Distribution of propensity scores as computed from our
example data set via logistic regression (solid black line) and
the respective fitted gaussian distributions (red dotted line) for
control (left) and treatment group (right).} \label{grafik11}
\end{center}
\end{figure}

We varied sample sizes between $100$ and $10,000$ with a step size
of $100$ and generated $5,000$ data sets for each single sample
size. Theis sample size range was informed by a random sample of
50 PS analyses, published in 2018, from the PUBMED database. For
each dataset, we generated standard normally distributed
observations, and randomly split the observations in control and
treatment groups. This simulation algorithm generates $i.i.d.$
observations, that is, balanced observations in both groups. Each
observation was assigned a random weight, which we wanted to
generate as realistic as possible. To this task, we used the
observed propensity scores from the example data set, which is
described in detail in the next section. We fitted Gaussian
distributions through the observed propensity scores in treatment
and control group (see figure \ref{grafik11}) and achieved means
of $0.30$ and $0.34$ and a identical standard deviation of $0.09$.
The propensity scores used in the simulation were then sampled
from these distributions, whereby we set values smaller than
$0.01$ or larger than $0.99$ to $0.01$ and $0.99$ respectively.
The samples propensity scores were then transformed to yield the
respective weights. In the first setting we used matching weights
to create rather similar weights with a small variability.

\begin{figure}
\begin{center}
\includegraphics[scale=0.4]{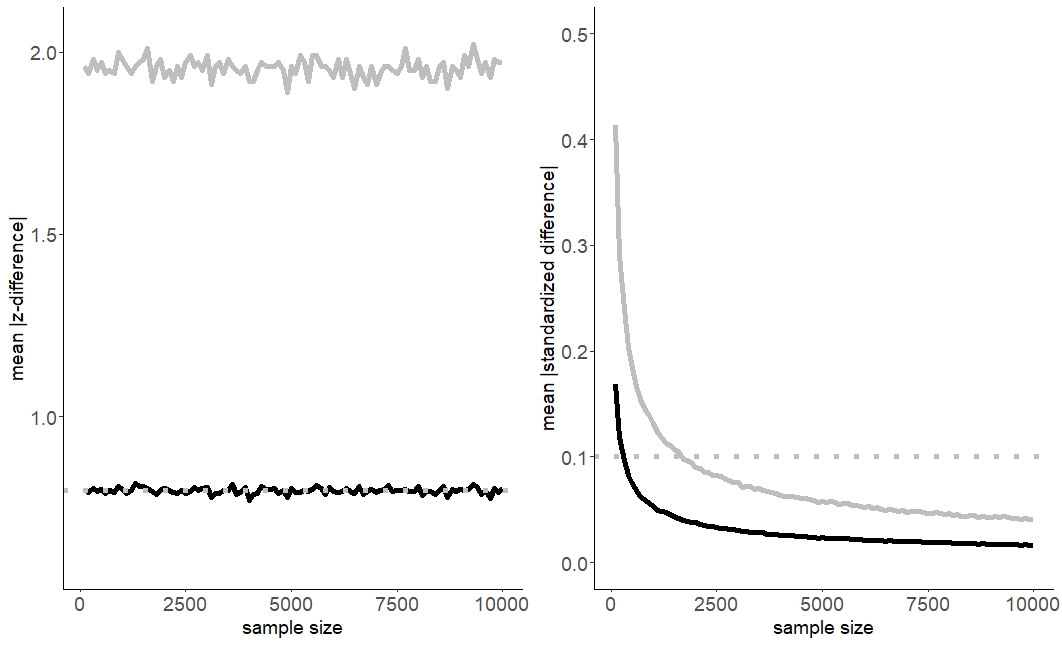}
\caption{Mean absolute value (solid line) and $95\%$ quantile
(dotted line) of the weighted z-difference (left) and weighted
standardized-difference (right). The grey dotted line represents
the expected value under randomization for the weighted
z-difference (left) and the value of $0.1$ (right), which is
commonly used to indicate good balance for the weighted
standardized difference.} \label{grafik3}
\end{center}
\end{figure}

In figure \ref{grafik3} we give mean absolute values for the
weighted z-difference and the weighted standardized difference,
respectively, across the different sample sizes. The value of the
weighted standardized difference depends on the respective sample
size with values considerably above $0.1$ for smaller sample
sizes. Actually, weighted standardized differences below $0.1$ are
in generally considered to indicate good balance
(\cite{normand2001validating}). However, with small sample sizes,
even good balance gives considerably larger values of the weighted
standardized difference and the rule-of-thumb of using $0.1$ as a
cut-off for good balance is too strict here.

Compared to this the weighted z-difference scatters narrowly for
the complete range of sample sizes around the value of
$\sqrt{\dfrac{2}{\pi}} \approx 0.79788$. This is the value that we
expect for the absolute value of a standard Gaussian distributed
random variable (\cite{bib19}).

\subsection{Continuous covariate, varying sample sizes, different distribution of weights}
\begin{figure}
\begin{center}
\includegraphics[scale=.3]{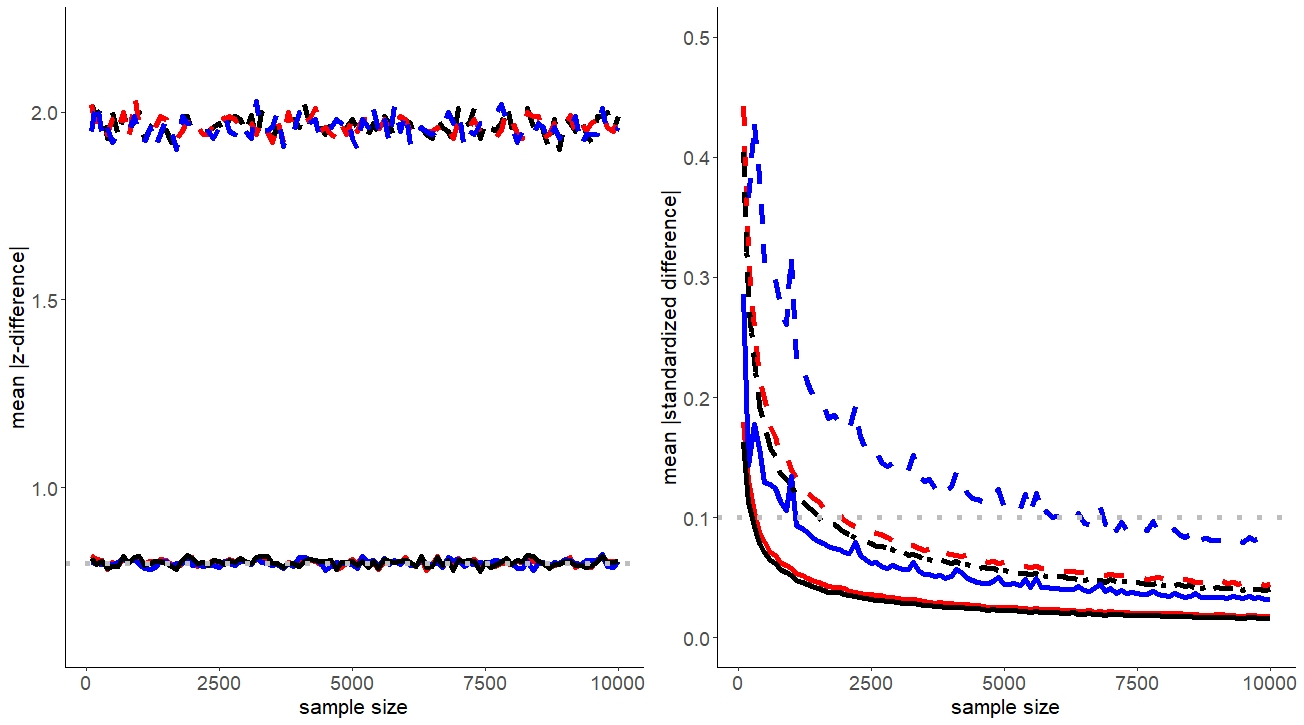}
\caption{Mean absolute value (solid line) and $95\%$ quantile
(dotted line) of the weighted z-difference (left) and weighted
standardized-difference (right) for different weighting methods
(black: unweighted, red: matching weights, blue: IPTW). The grey
dotted line represents the expected value under randomization for
the weighted z-difference (left) and the value of $0.1$ (right),
which is commonly used to indicate good balance for the weighted
standardized difference.} \label{grafik5}
\end{center}
\end{figure}

In the second simulation we changed the setting from the first
simulation by two aspects. First, we used three different
weighting schemes, which were (i) constant ($\equiv 1$) weights,
(ii) matching weights as in the previous simulation, and (iii)
IPTW weights. Second, we increased the variability of the weights
by doubling the variance of the distribution used for propensity
score generation. Because it is well known that IPTW weights have
increased variability as compared to other weighting methods, in
this second simulation we are able to assess the difference
between the weighted standardized difference and weighted
z-difference for weights with (i) no variability, (ii) small
variability and (iii) large variability. In figure \ref{grafik5}
we give the respective results. It is obvious that the
distribution of the weighted standardized difference depends on
the weights distribution, e.g., for a small sample size of $50$
observations for each group, we observe a mean value of $0.17$ for
constant weights, $0.19$ for matching weights, and above $0.3$ for
IPTW weights. This is again in conflict with the rule-of-thumb of
using the value of $0.1$ to separate good vs. bad covariate
balance. Opposed to this, the weighted z-difference does not
depend on the distribution of weights and scatters again around
its expected value of $\sqrt{\dfrac{2}{\pi}}$.

\subsection{Binary covariate, varying ratio of observations in treatment vs. control group}
In the third simulation we simulated the behaviour of the weighted
z-difference and the weighted standardized difference by changing
the ratio of sample size in control and treatment group. We chose
the case of a binary covariate and varied the percentage of
observations in the control group from $1 \%$ to $99 \%$. We kept
the overall sample size constant at $N=5,000$ for each single
percentage to have at least $50$ observations in each group. For
generating weights we used the Gaussian distributions and matching
weights as described for the first simulation setting.

\begin{figure}
\begin{center}
\includegraphics[scale=0.4]{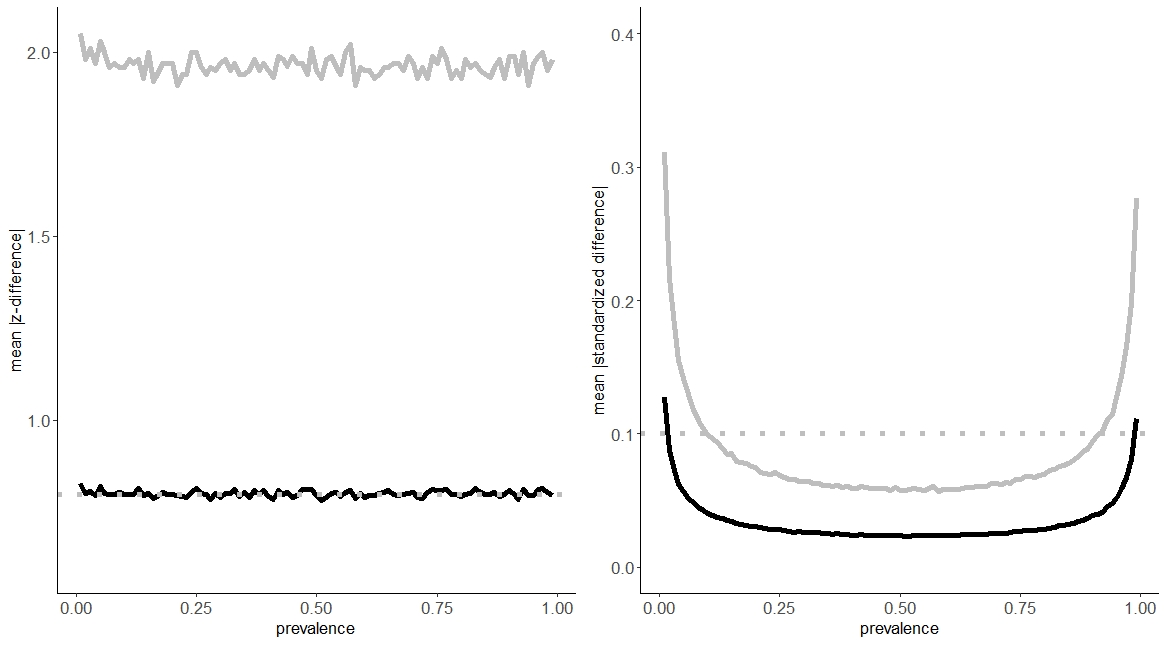}
\caption{Mean absolute value (black) and $95\%$ quantile (grey) of
the $5000$ replications of the weighted z-difference and the
weighted standardized difference (right) for a continuous variable
with prevalences ranging from $0.01$ to $0.99$. The grey dotted
line represents the expected value under randomization for the
weighted z-difference (left) and the value of $0.1$ (right), which
is commonly used to indicate good balance for the weighted
standardized difference.} \label{grafik6}
\end{center}
\end{figure}

In figure \ref{grafik6} we can see that the mean value of the
absolute weighted standardized difference is changing with the
different ratios. In fact we observe mean absolute values $>0.1$
if the sample size in one group is extremely smaller than in the
other (above $98\%$) and values $<0.05$ in the range between $8\%$
and $92\%$. In figure \ref{grafik6}, it can be seen that the
weighted z-difference is not affected by the group size ratio.

As a conclusion from the three simulation experiments we find that
the weighted standardized difference will only be of help for
assessing balance when sample size, weights distribution, and the
ratio of numbers in treatment and control group are simultaneously
taken into consideration. This is not the case for the weighted
z-difference which can be used to measure covariate balance
independent from these three factors.

\section{An applied example}\label{real_example}
In this chapter we apply weighted z-differences to data of a
published PS analysis in coronary bypass surgery (\cite{RN1001}).
The data set contains survival data of $1,282$ patients that
received either conventional coronary artery bypass grafting
($69.2\%)$ or a clampless off-pump coronary artery bypass
($30.8\%$) for the treatment of coronary artery disease. All
operations were performed at the Herz- und Diabeteszentrum NRW,
Bad Oeynhausen, Germany, between July $2009$ and November $2010$
and the decision, which operation was conducted was made by the
patient's surgeon. The covariates in table \ref{table1} were used
to estimate the propensity score by standard logistic regression
and to derive IPTW as well as matching
weights (\cite{li2013weighting}). Covariate balance was assessed by
using the weighted z-differences in the three differentially
weighted samples where all weights in the original, unweighted
data were set to one. Additionally we also give means and standard
deviation for the continuous covariates and relative frequencies
for binary, ordinal and nominal covariates in the weighted
samples. As the original data set does not contain a nominal
covariate, we randomly split the "yes" category of the covariate
variable "main stem stenosis" in two new categories resulting in a
now three-valued nominal covariate. In the unweighted sample, the
weighted z-differences reproduce, as expected, the standard
z-differences from the data set before PS-matching as given in our
previous publication (\cite{RN993}). We observe considerable
imbalance for age ($-3.24$), Diabetes ($3.39$), Priority ($-4.82$)
and the variance of LVEF ($2.82$), keeping in mind that an
absolute value of more than $1.96$ would indicate statistically
significant deviations from the balance that would be achieved by
randomization.

\begin{figure}
\begin{center}
\includegraphics[scale=0.4]{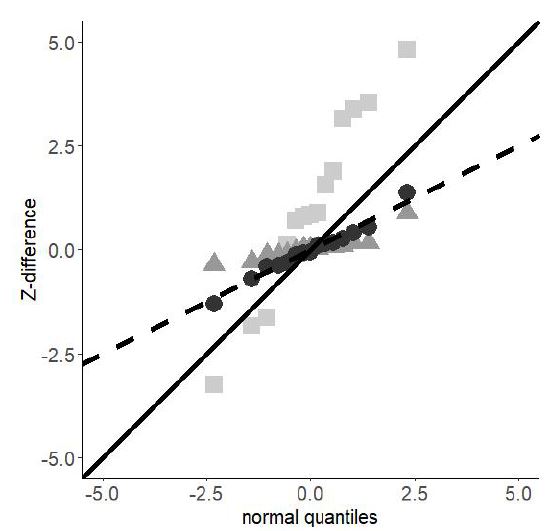}
\caption{Q-Q-plot of the weighted z-differences for the $15$
covariates in the example data set as given from three different
methods: raw data (square), matching weights (triangle) and IPTW
(circle). The solid black line and dotted black line indicate the
quantiles expected for a randomized trial and a perfectly
PS-matched data set.} \label{grafik7}
\end{center}
\end{figure}

\begin{table}[htbp]
\begin{tiny}
\caption{Covariate balance in the example data set as given for
unweighted (=raw) data, data weighted by matching weights, and
data weighted by IPTW.}\label{table1}

    \begin{tabular}{llllllllll}
    & \multicolumn{3}{c}{\textbf{Unweighted}} & \multicolumn{3}{c}{\textbf{Matching weights}} & \multicolumn{3}{c}{\textbf{Inverse probability weighting}} \\

     \textbf{Covariates} & \textbf{cCABG} & \textbf{Clampless OPCAB} & \textbf{Z-difference} & \textbf{cCABG} & \textbf{Clampless OPCAB} & \textbf{Z-difference} & \textbf{cCABG} & \multicolumn{1}{l}{\textbf{Clampless OPCAB}} & \textbf{Z-difference} \\

    \multicolumn{10}{c}{\textbf{Continuous scale (based on the difference of means)}}  \\
    Age (years) & 67.51 (9.44) & 69.31 (9.09) & -3.24 & 69.15 (8.88) & 69.31 (9.09) & -0.35 & 68.00 (9.29) & 67.15 (10.11) & 1.38 \\
    BMI (kg/m2) & 28.27 (4.47) & 27.80 (4.21) & 1.83  & 27.87 (4.14) & 27.80 (4.21) & 0.29  & 28.16 (4.37) & 28.31 (4.51) & -0.53 \\
    LVEF (\%) & 55.39 (14.06) & 56.66 (12.25) & -1.64 & 56.67 (13.47) & 56.66 (12.25) & 0.04  & 55.77 (13.90) & 56.04 (12.37) & -0.31 \\
    Previous surgeries (n) & 0.08 (0.39) & 0.05 (0.26) & 1.56  & 0.05 (0.26) & 0.05 (0.26) & 0.17  & 0.07 (0.35) & 0.09 (0.38) & -1.3 \\

    \multicolumn{10}{c}{\textbf{Binary scale}}  \\
    Gender (\% female) & 22.1  & 21.77 & 0.13  & 21.52 & 21.80 & -0.11 & 21.93 & 22.1  & -0.06 \\
    Hypertension (\%) & 84.1  & 82.28 & 0.8   & 82.85 & 82.45 & 0.17  & 83.69 & 84.71 & -0.4 \\
    Diabetes (\%) & 31.68 & 22.78 & 3.39  & 23.01 & 22.93 & 0.03  & 29    & 28.5  & 0.17 \\
    COPD (\%) & 7.1   & 5.82  & 0.88  & 5.64  & 5.85  & -0.13 & 6.65  & 5.99  & 0.41 \\
    Renal insufficiency (\%) & 1.24  & 0.76  & 0.84  & 0.82  & 0.76  & 0.1   & 1.11  & 1.02  & 0.15 \\
    Previous MI (\%)  & 35.74 & 27.09 & 3.14  & 27.36 & 27.26 & 0.03  & 33.14 & 35.29 & -0.71 \\
    Previous stroke (\%) & 2.37  & 1.01  & 1.89  & 1.07  & 1.02  & 0.07  & 1.96  & 1.85  & 0.11 \\
    PAD (\%) & 11.39 & 11.9  & -0.26 & 11.47 & 11.74 & -0.13 & 11.4  & 10.83 & -0.27 \\
    Preoperative IABP (\%) & 1.47  & 1.01  & 0.7   & 1.07  & 1.02  & 0.08  & 1.34  & 1.4   & -0.08 \\

    \multicolumn{10}{c}{\textbf{Ordinal scale}}  \\
    Priority (\%) &       &       & 4.82  &       &       & -0.88 &       &       & -0.38 \\
    Elective       & 80.95 & 91.90 &       & 89.39 & 91.85 &       & 83.55 & 83.53 &  \\
    Urgent        &  9.81 &  2.53 &       & 6.58  & 2.55  &       & 8.81  & 2.89  &  \\
    Emergent    &  8.68 &  5.32 &       & 3.81  & 5.35  &       & 7.18  & 13.04 &  \\
    Ultima ratio &  0.56 &  0.25 &       & 0.22  & 0.25  &       & 0.46  & 0.54  &  \\

    \multicolumn{10}{c}{\textbf{Nominal scale}}  \\
    Main stem stenosis (\%) &       &       & -3.53 &       &       & -0.03 &       &       & 0.11 \\
    Yes  & 25.32 & 25.48 &       & 25.46 & 25.54 &       & 24.73 & 25.47 &  \\
    No   & 33.16 & 23    &       & 32.76 & 32.12 &       & 25.22 & 25.87 &  \\
    Unclear & 41.52 & 51.52 &       & 41.78 & 42.34 &       & 50.05 & 48.66 &  \\

    \multicolumn{10}{c}{\textbf{Continuous scale (based on the difference of variances)}}  \\
    Age (years) & 67.51 (9.44) & 69.31 (9.09) & 0.9   & 69.06 (8.82) & 69.26 (9.09) & -0.65 & 68.00 (9.29) & 67.15 (10.11) & -2.02 \\
    BMI (kg/m2) & 28.27 (4.47) & 27.80 (4.21) & 1.08  & 27.90 (4.14) & 27.82 (4.21) & -0.27 & 28.16 (4.37) & 28.31 (4.51) & -0.54 \\
    LVEF  & 55.39 (14.06) & 56.66 (12.25) & 2.8   & 56.62 (13.52) & 56.66 (12.25) & 1.83  & 55.77 (13.90) & 56.04 (12.37) & 2.13 \\
    Previous surgeries (n) & 0.08 (0.39) & 0.05 (0.26) & 1.09  & 0.05 (0.26) & 0.05 (0.26) & 0.02  & 0.07 (0.35) & 0.09 (0.38) & -0.28 \\

    \end{tabular}
\end{tiny}
\end{table}

Covariate balance is largely improved in the weighted samples
after applying IPTW and matching weights, where the matching
weights method outperforms the standard IPTW method: the mean
absolute value of the weighted z-differences is $1.64$ (variance
$1.67$) for equal weights, $0.59$ ($0.33$) for IPTW, and $0.24$
($0.18$) for matching weights. The different performances in
covariate balancing can also be seen from the Q-Q-plot for the
weighted z-differences in figure \ref{grafik7}. From such a plot
we would expect to see (unweighted) z-differences from a
randomized trial to lie on a line through the origin with a unit
slope (\cite{RN993}). In addition, unweighted and weighted
z-differences could also be compared to a perfectly matched PS
analysis in the sense of \cite{RN1002,RN1003} from
which we would expect the values to lie on a line through the
origin with slope $0.5$. The raw unweighted z-differences clearly
deviate from the line with slope $1$, indicating systematic
differences between the original treatment groups. However, for
both weighting methods the slope of the weighted quantiles is
smaller than $0.5$, pointing to an even better balance in the
weighted samples as compared to a perfectly matched sample.

\section{Discussion}\label{discussion}
In this paper, we propose weighted z-differences to measure
covariate balance in weighted propensity score analyses. In
comparison to the weighted standardized difference of Austin, we
see three advantages of weighted z-differences. First, weighted
z-differences can be used for continuous, binary, ordinal and
nominal covariates, and not just for continuous and binary ones.
Second, the distribution of the weighted z-difference is
independent of the sample size, which allows comparison of data
sets of different sizes. Third, the distribution of the weighted
z-difference is, again in contrast to the weighted standardized
difference, independent of the distribution of the weights.
Weighted z-differences are straightforward generalizations of our
previous proposal of z-differences in matched PS analyses and, at
least in the continuous and the binary covariate case, reduce to
standard z-differences in the unweighted case. The idea of the
double transformation from a $\chi^2$ to a standard normal
distribution in the nominal covariate case now also allows for a
standard, unweighted z-difference. Finally, weighted z-differences
can be used for all types of weights (e.g., standard IPTW
weights (\cite{RN1016}), but also matching (\cite{li2013weighting}),
optimal (\cite{RN1025}), or ATT/ATC weights (\cite{RN1025})), and thus
not only allow optimizing and comparing covariate balance with
respect to the covariates included, but also with respect to the
respective weight type. Due to the "z-property" of weighted and
unweighted z-differences, covariate balance can also be compared
(e.g., by using the sum of z-differences across all covariates)
between a weighted PS analyses and PS-matching. As such, also the
previously proposed Q-Q-plots to compare covariate balance 1)
across PS methods, 2) to the expected balance in a randomized
trial, and 3) to a perfectly PS matched data set in the sense of
Rubin and Thomas can be used. We also emphasize that the
z-difference is a global balance measure that can also be used
outside of propensity score methods, e.g. in randomized trials, to
measure the balance of observations in two groups.

It is fair to point to some limitations of weighted z-differences.
We observed the weighted z-difference for nominal covariates to be
somewhat sensitive against very low numbers of observations in
some categories of the covariate, and Gagunashvili recommended to
have at least ten observations within each cell (\cite{RN1004}). In
the case of very sparse data, we propose as a solution to
calculate weighted binary z-differences for each dichotomization
of the nominal covariate against a reference category. Another
limitation of the weighted z-difference in the PS setting is that
it can be only applied with PS-weighting methods. In contrast,
\cite{bib5} showed, how weighted standardized differences
can also be used when stratifying on or adjusting for the
propensity score.

Balancing advantages and limitations we recommend weighted
z-differences for assessing covariate balance in weighted PS
analyses.

\medskip

\noindent {\bf{Conflict of Interest}}

\noindent {\it{The authors have declared no conflict of interest.
(or please state any conflicts of interest)}}

\end{document}